# Exchange bias and training effect in an amorphous zinc ferrite/ nanocrystalline gallium ferrite bilayer thin film

Himadri Roy Dakua

Department of Physics, Indian Institute of Technology Bombay, Mumbai, India - 400076

Email: hroy89d@gmail.com

## Abstract

In this paper I report, exchange bias effect in a bilayer thin film of amorphous zinc ferrite and nanocrystalline gallium ferrite. The amorphous zinc ferrite layer was deposited at room temperature ($T_S$ = RT) on top of a nanocrystalline gallium ferrite thin film using a pulsed laser. This bilayer film showed large exchange bias effect ($H_E$ ~ 418 Oe at 2 K). The exchange bias shift decreased exponentially as the temperature increased and disappeared for T > 30 K. Along with the exchange bias shift the film also showed enhanced magnetization in Field Cooled (FC) measurements as compared to the Zero Field Cooled (ZFC) magnetization. The bilayer film also showed training effect at 2 K, which followed spin configurational relaxation model. The observed exchange bias effect could be attributed to the pinning anisotropy of the spin glass amorphous zinc ferrite layer pinned at the interface of gallium ferrite.

## 1. Introduction

The coupling between two different magnetic materials along their interface provides many phenomena of both scientific and technological importance[1-3] such as the exchange bias effect[4-6], proximity effect[7-9], exchange spring permanent magnet[10-12] etc. The coupling between an Antiferromagnetic (AFM) and a Ferromagnetic (FM) materials generally provides the exchange bias effect.[5,13] The exchange bias effect manifests a shift in the magnetic hysteresis loop along the field axis when the system is field cooled below the Néel temperature of the AFM phase.[5,6] The utility of the exchange bias systems in many magnetoelectronics devices[14-16] helped to grow the research interest in this topic and it also led to the discovery of exchange bias effect in a lot of different systems such as AFM)- FM[17-19], AFM-Ferrimagnetic (FIM)[20-22], FIM-FM[23], Spin Glass (SG)-FM[24] heterostructure oxides, alloys and nanomaterials[25] and also in some single phase (crystallographic) materials.[25,26] However, the complete understanding and a universal microscopic model for the exchange bias effect are yet to be achieved.

The earliest model for the exchange bias effect was provided by Meicklejohn and Bean and the exchange bias field was expressed as[27-29]

$$H_E = -\frac{JS_{FM}S_{AFM}}{t_{FM}M_{FM}}. \qquad (1)$$

Where J is the exchange coupling constant between the interfacial magnetization $S_{AFM}$ and $S_{FM}$ of the AFM and FM layers respectively. $M_{FM}$ and $t_{FM}$ are the saturation magnetization and the thickness of the FM layer respectively.[30] The equation 1 indicates that the interfacial magnetization of AFM layer plays an important role for the exchange bias effect, as the saturation magnetization of the FM layer ($M_{FM}$) and $S_{FM}$ are most likely to remain unchanged.[30] The uncompensated AFM spin states at the interface provide a net AFM interfacial moment, $S_{AFM}$, along (or opposite to) the field direction. These uncompensated spins are in thermodynamic non-equilibrium states as compared to the compensated AFM spins. The decay of this interfacial moment towards an equilibrium state leads to a decrease in the exchange bias effect in the training measurements or as the temperature increased.[28,30]

In this paper, I show the exchange bias and training effect in a bilayer thin film of amorphous zinc ferrite and nanocrystalline gallium ferrite. Gallium ferrite is one the few materials that show near room temperature magnetoelectric properties.[31,32] The coupling between the magnetic and ferroelectric/ dielectric properties of a magnetoelectric material provides an opportunity to control the one by means of the other.[33,34] The exchange bias effect in these magnetoelectric systems, might help to control the exchange bias related phenomena by an external electric field.[3] This unique property of the system might be useful in future magnetic devices. The observed exchange bias effect in this amorphous zinc ferrite and nanocrystalline gallium ferrite bilayer thin film is found to obey the Meicklejohn and Bean model (equation 1) as the interfacial magnetization of the AFM layer is replaced by that of the spin glass amorphous zinc ferrite layer.

## 2. Experimental Details

The gallium ferrite (GFO) thin film was deposited on an amorphous quartz substrate using a Nd:YAG pulsed laser. A single phase high density GFO PLD target was used for the deposition. In this process, the laser energy density was kept at ~2 Joule/ cm$^2$, pulse repetitions rate was 10 shots/sec and the deposition duration was 30 minutes. The film was deposited in oxygen atmosphere (0.16 mbar pressure) and the substrate was kept at room temperature ($T_S$ = RT) at 4.5 cm from the PLD target. The deposited single layer thin film was ex-situ annealed at $T_a$ = 750°C for 2 hours, in air atmosphere. This annealed film showed single phase nanostructured gallium ferrite features. On top of this nanostructured gallium ferrite thin film an amorphous zinc ferrite layer was deposited. A stoichiometric, high density zinc ferrite (ZFO) PLD target was used to deposit this layer. The deposition conditions for the top amorphous zinc ferrite layer was kept same as that of the single layer gallium ferrite. The as deposited bilayer film was used to study the EB effect.

### 3. Results and discussions
#### a. Structural and microstructural properties of the film

The structural and the microstructural properties of the bilayer film were studied using XRD and FEG-SEM. Fig. 1 shows the XRD of the bilayer thin film deposited on the amorphous quartz substrate. The XRD of the bilayer thin film is compared with the XRD of single layer gallium ferrite thin film (red data) and the single layer zinc ferrite thin film (black data). The single layer GFO thin film was deposited and annealed in same conditions as that of gallium ferrite layer of the bilayer thin film. Similarly, ZFO single layer thin film was also deposited in same conditions as that of the top layer of the bilayer thin film. The XRD of the single layer ZFO thin film deposited on amorphous quartz substrate at $T_S$ = RT did not show any Bragg's peaks which indicates that the deposited film is amorphous (within the limit of the XRD). The amorphous nature of ferrite film deposited (using Physical Vapour Deposition (PVD) process) at room temperature (RT) is a commonly observed phenomenon.[35-37] On the other hand, the GFO single layer film deposited at $T_S$ = RT and annealed at $T_a$ = 750 °C for 2 hours shows Bragg's peaks corresponding to the orthorhombic crystal structure of GaFeO$_3$ of space group Pc2$_1$n. While the bilayer film shows fewer Bragg's peaks which are observed at same positions as that of the single layer GFO film. However, similar to the ZFO single layer amorphous thin film, the bilayer film also did not show any peak either corresponds to the Zn-ferrite cubic spinel phase or any other impurity phase. This indicates that the top layer of the bilayer film is amorphous zinc ferrite.

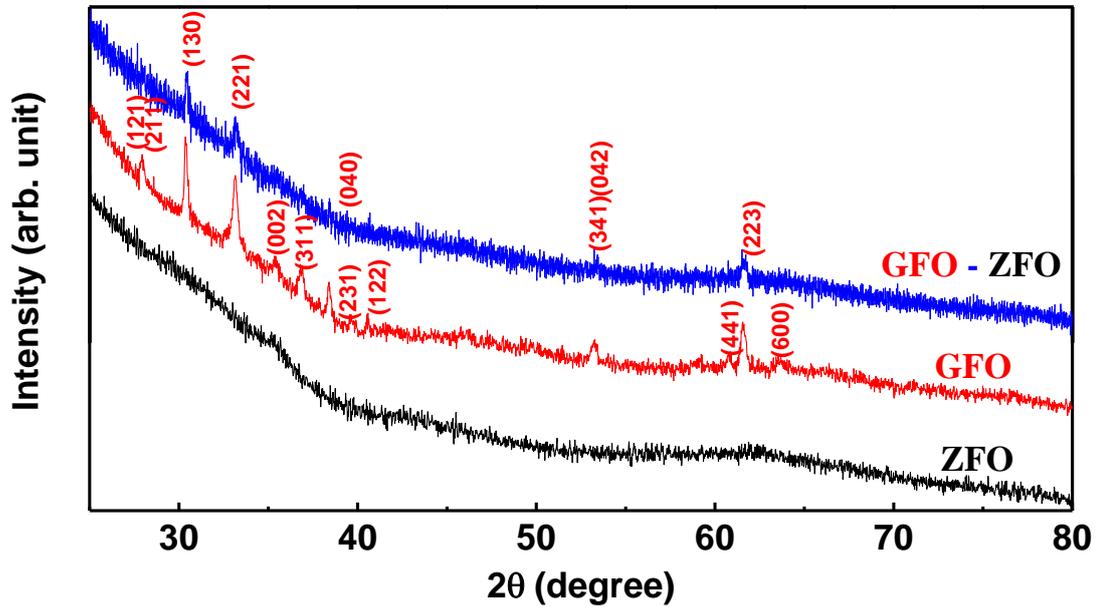

Fig. 1. XRD of the samples. GFO-ZFO: XRD of the amorphous zinc ferrite/ nanocrystalline gallium ferrite bilayer, GFO: XRD of the nanocrystalline gallium ferrite single layer and ZFO: XRD of the amorphous zinc ferrite single layer.

The FEG-SEM of the single layer GFO thin film deposited at $T_S$ = RT and annealed at $T_a$ = 750 °C for 2 hours are shown in Fig. 2 (a - b). This film showed almost isolated nanostructures of average size ~100 nm. The thickness of the single layer nanostructured GFO film is ~160 nm. Fig. 2 (c) shows the FEG-SEM image of the bilayer thin film. Fig. 2 (d) shows the cross sectional FEG-SEM image of the bilayer thin film. The thickness of the bilayer film is ~ 275 nm. Therefore, the average thickness of the top amorphous ZFO layer could be estimated as ~115 nm. However, a rough interface of these two layers could be expected as the bottom gallium ferrite layer showed isolated nanostructure like features.

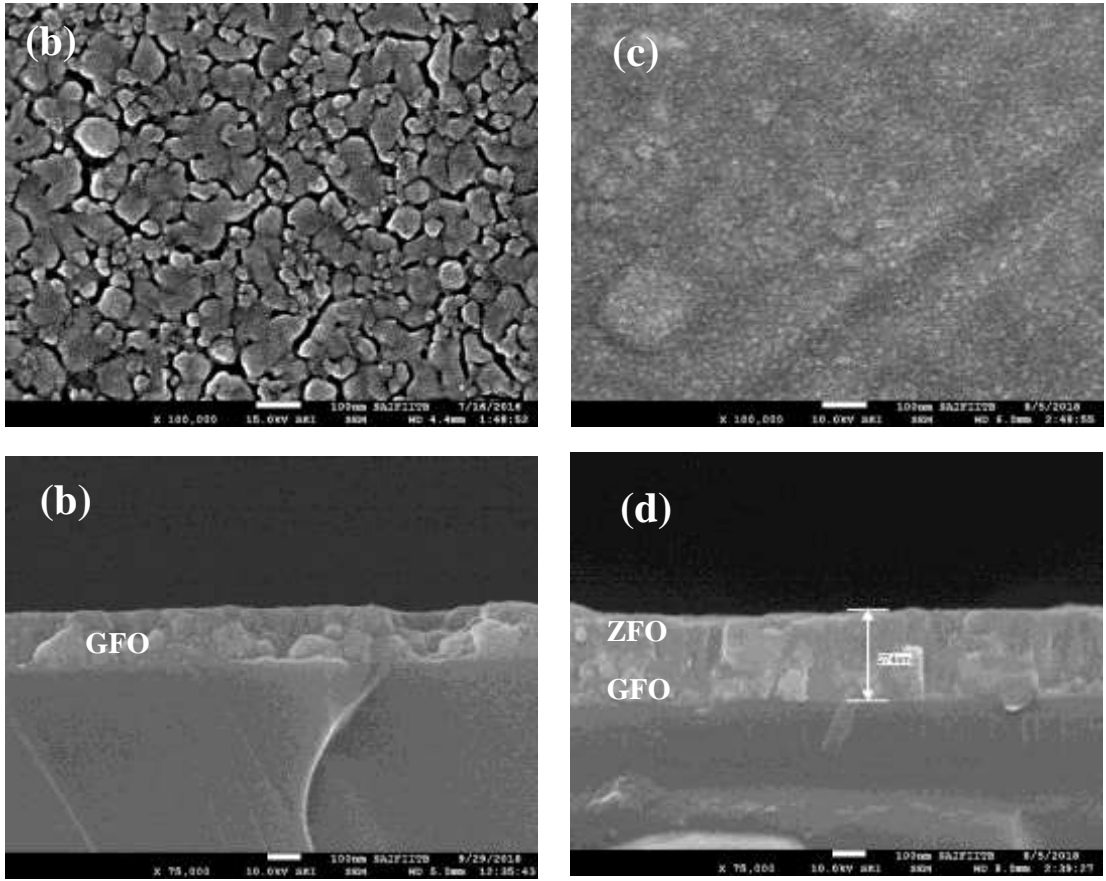

Fig. 2. FEG-SEM images of the single gallium ferrite and gallium ferrite/ amorphous zinc ferrite bilayer. (a) Top surface of single layer GFO film, (b) cross-section of single layer GFO film, (c) top surface of bilayer film and (d) cross-section of the bilayer film.

### b. Magnetic properties of the film

Fig. 3 shows the M-T data of the bilayer film measured in both Field Cooled (FC) and Zero Field Cooled (ZFC) mode. The magnetic field was applied along the film's plane. The inset of the figure shows derivative of the FC magnetization, dM/dT as a function of temperature. The transition temperature ($T_C$) was obtained from the minimum of this data ($T_C$ = 235 K), which is comparable with the $T_C$ of the single layer nanostructured gallium ferrite thin film and bulk gallium ferrite polycrystalline sample.[31,32] A large difference in the FC and the ZFC magnetization of the film could be due to the high anisotropy of gallium ferrite[38,39] and spin glass properties of top amorphous layer.

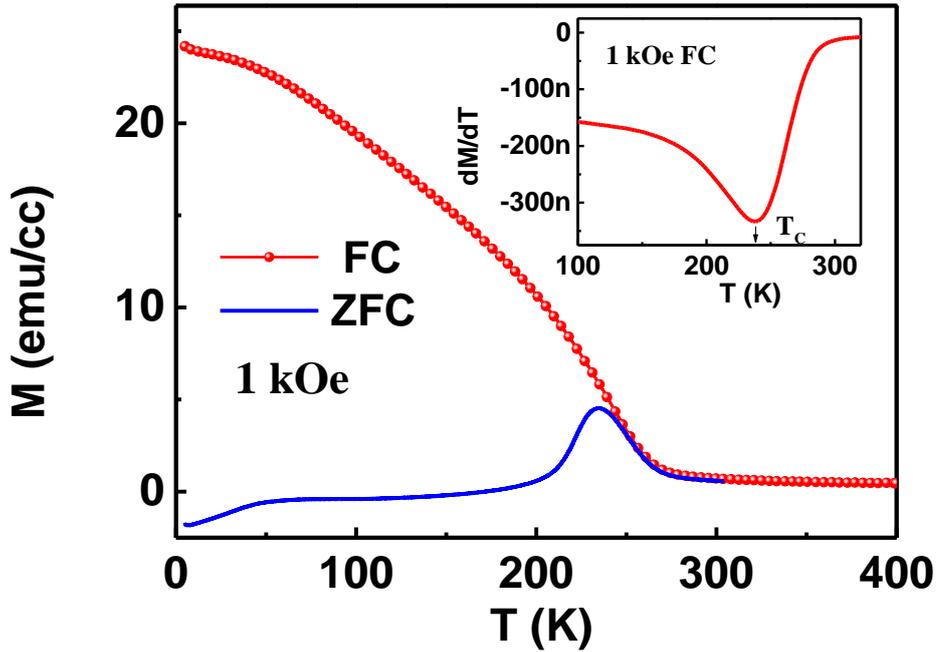

Fig. 3. Temperature dependence ZFC and FC magnetization of the amorphous ZFO/ nanocrystalline GFO bilayer film. The ZFO layer was deposited on top of nanocrystalline GFO layer, at $T_S$ = RT.

Fig. 4 shows a part of the 300 K M-H curve and 5 K ZFC M-H loop of the bilayer thin film. As expected from the M-T data, the 300 K M-H data shows almost linear variation with the magnetic field and no coercivity. This clearly indicate that the film is in paramagnetic state at room temperature. However, the 5 K ZFC M-H loop shows a large coercivity ($H_C = (H_{C2} - H_{C1})/2$ = 5560 Oe), which is equivalent to the coercivity of the single layer nanostructured gallium ferrite thin film. The 5 K ZFC M-H loop also shows unsaturated behaviour. However, unlike the minor M-H loops, this data did not show any vertical shift of the loop ($|M_{+50\ kOe}| = |M_{50\ kOe}|$) and it also showed almost reversible behaviour in the high field ZFC M-H curves. The unsaturated behaviour of the M-H loop could be attributed to the high anisotropy of gallium ferrite and the glassy behaviour of the top amorphous zinc ferrite layer.[39,40]

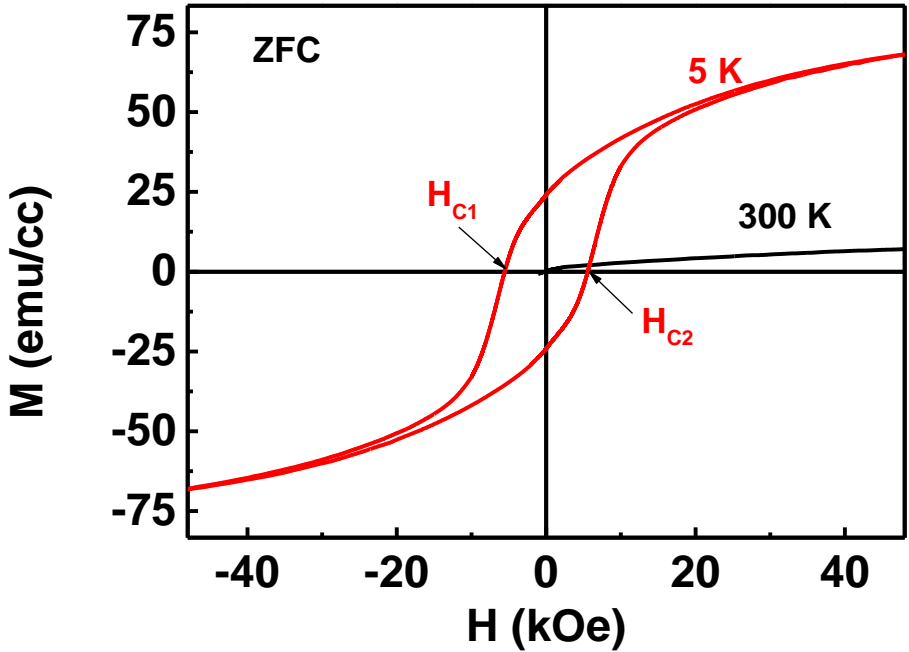

Fig. 4. ZFC M-H data of the film measured at 5 K and 300 K

### c. Exchange bias effect in the film

This amorphous ZFO/ nanocrystalline GFO bilayer thin film showed exchange bias effect. Here, I present the details of the observed exchange bias effect in this film. Fig. 5 shows zoomed view of a ZFC and a FC M-H loops measured at 2 K. For the FC measurement, the sample was cooled down to 2 K from RT in presence of 50 kOe magnetic field applied along the film's plane. The FC M-H loop shows exchange bias shift ($H_E = (H_{C2} + H_{C1})/2$) along the negative field axis. The magnitude of exchange bias shift at 2 K is $H_E = 418$ Oe. The coercivity of the FC M-H loop is $H_C = 5938$ Oe at 2 K which is higher than the ZFC $H_C$ (= 5840 Oe at 2 K). The increase in coercivity ($H_C$) due field cooling is a commonly observed phenomena in different exchange bias systems.[6,41,42] The pinned spins of the AFM or glassy states at the interface of the ferromagnetic layer contribute significantly in the coercivity enhancement of the FC M-H loops. These pinned spins switched irreversibly while reversing the magnetization of the FM layer. The work done due to this irreversible switching resulted in broadening of the M-H loop.[41]

One also needs to note that the remanence magnetization ($M_{r1}$) of the film is increased in the FC M-H loop as compared to the ZFC M-H loop. This increase in remanence magnetization is also associated with an increase in the high field magnetization ($M_{50\,kOe}$) of the FC M-H loop as compared to the ZFC M-H loop. The enhancement of magnetization of the FC M-H loop as compare to the ZFC M-H loop was also reported in different EB systems. [43-45] This increased high filed magnetization of the bilayer film could be due to the pinning of the interface spins of the glassy amorphous ZFO layer along magnetization of the nanocrystalline gallium ferrite layer.

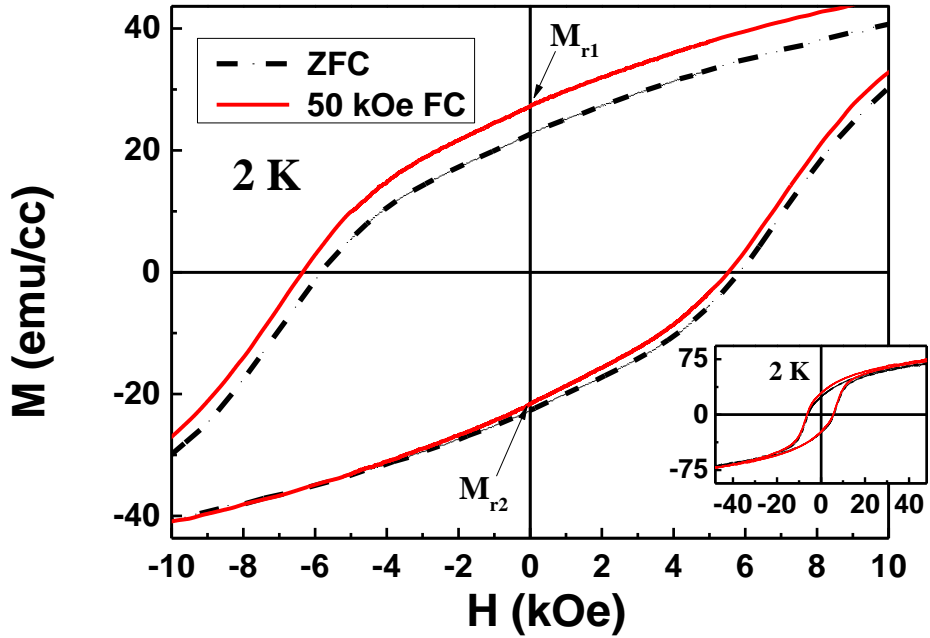

Fig. 5. Low field part of 2 K ZFC and FC M-H loops. The film shows exchange bias effect. Inset: the M-H loops are shown for ± 50 kOe field range.

The temperature dependence of the exchange bias effect is also studied in this film. Fig. 6 (a) shows the exchange bias field $H_E$ of 50 kOe FC M-H loops of the bilayer film measured at different temperatures. Fig. 6 (b) shows the coercivity $H_C$ of the FC M-H loops as a function of temperature. The exchange bias shift ($H_E$) decreased monotonically as the temperature increased and it almost disappeared at T > 30 K. The coercivity ($H_C$) of the sample also decreased as the temperature increased. These decrease in the exchange bias shift ($H_E$) and the coercivity ($H_C$) of the film followed an exponential decay as shown in equation 2 (a) and (b).

$$H_E = H_{E0} e^{-T/T_0} \qquad (2.a)$$

$$H_C = H_{C0} e^{-T/T_{c0}} \qquad (2.b)$$

Where $H_{E0}$ and $H_{C0}$ are the exchange bias field and coercivity at T = 0 K.

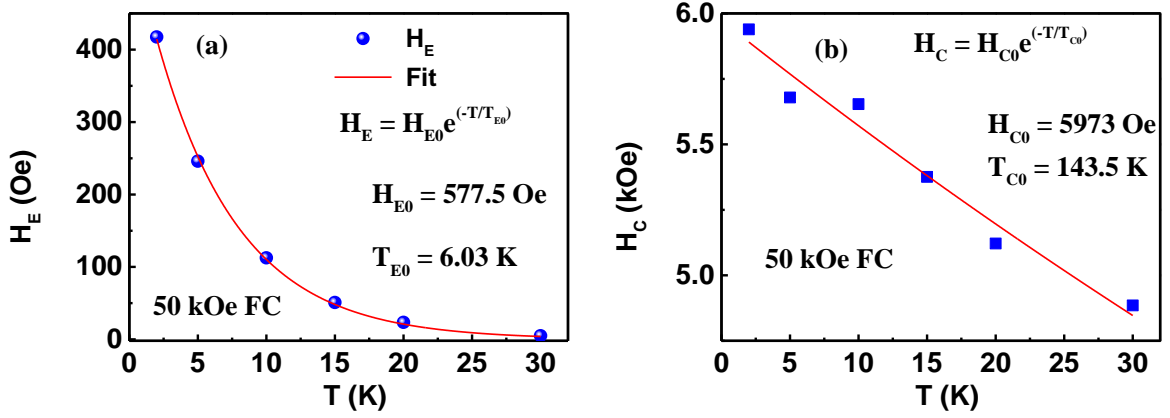

Fig. 6. The temperature dependence of the (a) exchange bias field $H_E$ and (b) coercivity $H_C$ of the film. The solid lines are the fitted data

The value of the fitted parameters are shown in the Fig. 6 (a-b). The exponential decrease of the exchange bias field ($H_E$) and the coercivity ($H_C$) were generally reported in the systems with spin glass or magnetically frustrated interfaces[46-48] such as $SrMnO_3$ and $La_{0.7}Sr_{0.3}MnO_3$ bilayer, where the competing magnetic order led to formation of a spin glass state at the interface.[49] The increase in temperature unfreeze the glassy spins of the amorphous zinc ferrite layer and turned them into paramagnetic spins above the blocking temperature. The spins in the paramagnetic region could not provide unidirectional anisotropy to the system and the exchange bias effect disappeared in higher temperature.

### d. Training effect

The exchange bias systems generally show a decrease in the exchange bias shift due to consecutive M-H loop iterations, which is known as training effect.[28,30,50,51] The training effect is also observed in this amorphous ZFO/ nanocrystalline GFO bilayer thin film. The training effect was measured after cooling down the film from room temperature to 2 K in presence of 50 kOe applied magnetic field. The field was cycled from +50 kOe to -50 kOe to +50 kOe for each consecutive M-H loop iterations. Fig. 7 (a) shows the low field part of training M-H loops of the film for 1st and 8th iterations (n = 1 and 8) along with their corresponding inverted loops, (-M)-(-H). Fig. 7 (b) shows the high field part of the n = 1 and 8 M-H loops.

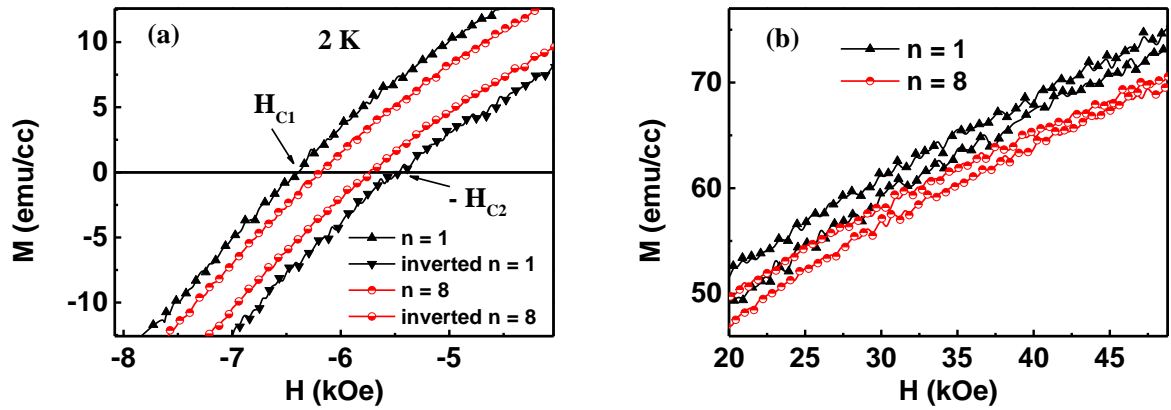

Fig. 7. Training effect of the film measured at 2 K after cooling in 50 kOe magnetic field. (a) Normal and inverted low field magnetization curve of first (n = 1) and n = 8$^{th}$ M-H loops iterations. (b) High field magnetization of the n = 1 and 8 M-H loops.

The exchange bias shift ($H_E$) as well as the high field magnetization ($M_{50\,kOe}$) of the film decreased as the loop iterations (loop index: 'n') increased. Fig. 8 shows the variation of the exchange bias field ($H_E$) obtained as a function of 'n' in this training measurements.

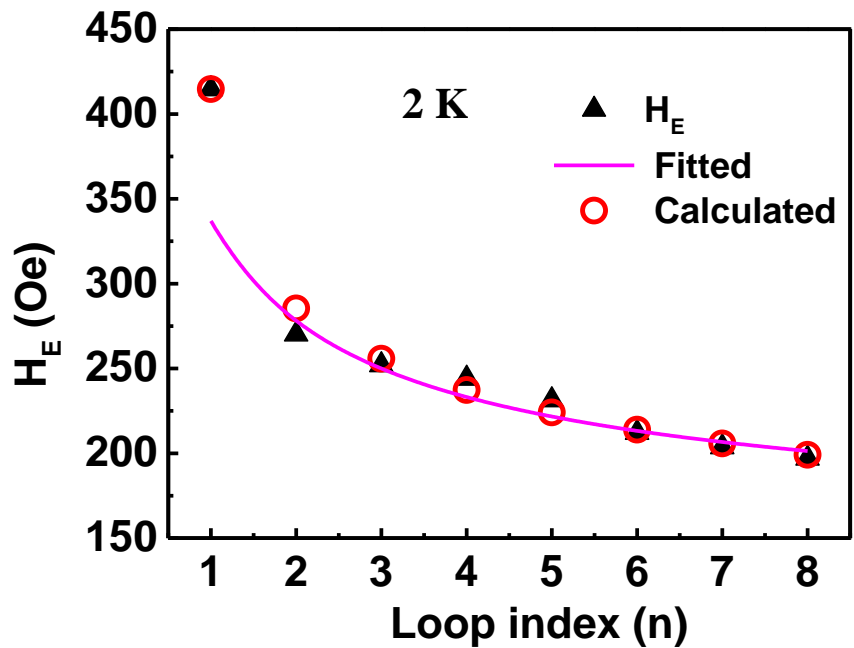

Fig. 8. Training loop index (n) dependence of the exchange bias field $H_E$.

The decrease in the $H_E$ of the film showed a very similar trend as compared to the training effect of different EB systems reported previously.[6,25] I fitted these exchange bias fields with the well-known empirical power law for training effect

$$H_E(n) - H_E^\infty = \frac{k}{\sqrt{n}} \qquad (3)$$

Where, $H_E^\infty$ is the exchange bias field for n = ∞ and k is a sample dependent proportionality constant.[52] The exchange bias field shows good fit with equation 3 for n > 1. The solid curve in the Fig. 8 represents the fitted data with this equation. The fitted data is extrapolated upto n = 1. The fitted parameters are k = 217.3 Oe and $H_E^\infty$ = 124.5 Oe. This empirical power law behaviour of the exchange shift was observed in many exchange bias systems.[6,25] However, the physical origin of such behaviour is still not very clear. Hochstrat *et al.* had also observed similar power law decay of the training exchange bias fields with n (for n > 1) in NiO/Fe heterostructure.[28] Later, Binek[30] proposed an analytical model for the exchange bias training effect in the same system. He considered thermodynamic non-equilibrium spin states at the interface of the AFM-FM heterostructure. The training measurements lead to spin configurational relaxation of the interface AFM spins towards an equilibrium state. The process was formulated as [30]

$$H_E(n+1) - H_E(n) = -\gamma(H_E(n) - H_E^\infty)^3 \qquad (4)$$

Here, γ is a sample dependent parameter. The experimentally observed $H_E(1)$ is considered as the initial value while calculating $H_E(n)$ using this recursive formula.[30] The circles in the Fig. 8 represents data obtained from this calculation. The parameters corresponding to these calculated data are $H_E^\infty$ = 81.5 Oe and γ = 3.494 ×10$^{-5}$ Oe$^{-2}$. We can see that a much closer to the experimental data was obtained using the equation 4.

The other important features of the exchange bias effect in this film are the enhancement of the high field magnetization in the FC M-H loop as compared to the ZFC M-H loop and the decrease in the high field magnetization with the increasing M-H loop iterations in the training measurements. Fig. 9 shows the exchange bias shift of the film as a function of average high field magnetization $M_{50\ kOe}$ $(= \frac{M_{+50\ kOe} + |M_{-50\ kOe}|}{2})$ obtained in the training measurements. The inset of Fig. 9 shows the $M_{50\ kOe}$ as a function of the training loop index 'n'.

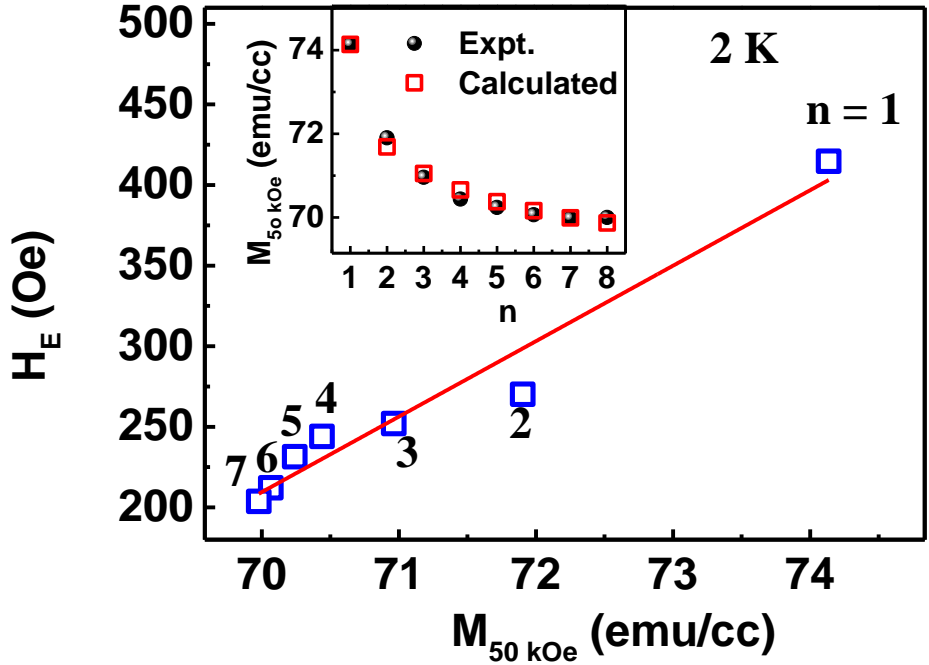

Fig. 9. Almost linear decrease in $H_E$ with the magnetization is observed in the training measurements of this system. Inset: $M_{50\,kOe}$ as a function of 'n'.

Similar to the exchange Bias field ($H_E$), the high field magnetization ($M_{50\,kOe}$) of the film also decreased with the increasing 'n'. According to Binek,[30] this decrease in the high field magnetization could be attributed to the relaxation of the coupled thermodynamically nonequilibrium interfacial spins of the amorphous ZFOlayer towards an equilibrium state as a result of consecutive M-H loop iterations. Thus the decrease of the high field moment could be formulated in a similar way as that of the exchange bias shift ($H_E$) in equation 4.[30]

$$M_{50\,kOe}(n+1) - M_{50\,kOe}(n) = -\gamma'(M_{50\,kOe}(n) - M_{50\,kOe}^{\infty})^3 \qquad (5)$$

The inset of Fig. 9 shows the calculated data (open squares) using the equation 5. Almost similar to the experimental data was obtained by using $M_{50\,kOe}^{\infty}$= 67.37 emu/cc and $\gamma' = 7.93\times10^{-3}$ (emu/cc)$^{-2}$ in the recursive equation 5. According to Meicklejohn and Bean model (equation 1)[27], the exchange bias shift should decrease linearly with the decrease in the coupled interface moment ($S_{AFM}$). Fig. 9 shows that our sample also shows almost a linear decrease in the exchange bias field with the decrease in the average high field moment of the sample.

It is known that the exchange bias effect requires two magnetic phases, a reversible magnetic phase coupled with another phase that fixes this reversible phase in a certain direction (along or opposite the field direction for negative and positive exchange bias effect respectively).[43] In this system, the amorphous ZFO layer act like the reversible phase. This amorphous zinc ferrite top layer of the bilyer film turned into a spin glass state at low temperature due to the presence of competing local exchange interactions.[40] The spin glass amorphous ZFO have multiple spin configurational ground states[43,53] and spins get frozen in these ground states in random directions below the blocking temperature. However, as the system was field cooled (from room temperature, which is much above the blocking temperature), some spins of the glassy amorphous ZFO layer get frozen along the applied magnetic field direction (or along magnetization of GFO layer) below the blocking temperature. The aligned frozen interfacial spins of the glassy layer pinned at the interface of the ferrimagnetic ($GaFeO_3$) layer and provided a pinning unidirectional anisotropy to it, which resulted in an exchange bias shift in FC M-H loop of the bilayer film. The spins pinned along the magnetization of GFO layer also increased the net magnetization of the bilayer film. However, it is likely that these aligned pinned spins of the interfacial glassy, amorphous ZFO layer are in higher energy states as compared to their ground state energy.[28,30,53] The consecutive M-H loop iterations helped to relaxed some of the aligned spins into the permissible ground states of the spin glass configurations. This led to a decrease in the magnetization as well as the exchange bias shift in the training effect measurements. Similarly, the temperature variation could be attributed to the decrease in the pinned spins of the glassy states as a result of increased thermal fluctuation with increasing temperature.

## 4. Conclusion

In this paper a detail study of exchange bias effect in an amorphous zinc ferrite/ nanocrystalline gallium ferrite bilayer thin film is presented. The exchange bias shift of the film was found to decrease exponentially as the temperature increased. The exchange bias effect was also associated with an increase of the net magnetization of the Field Cooled film as compared to the Zero Field Cooled. This enhancement in the magnetization is attributed to the interfacial magnetization of the spin glass amorphous zinc ferrite layer. The exchange bias shift was also found to decrease linearly with the decrease in the net magnetization (interfacial) of the film in the training effect measurements, which is a typical behaviour of the exchange effect one could expect from the Meicklejohn - Bean model.

## Acknowledgements

The author thanks Prof. Shiva Prasad and Prof. N. Venkataramani for the PLD facilities. The author thanks B. N. Sahu for providing the zinc ferrite PLD target. The author also thanks SAIF and IRCC of IIT Bombay, for the VSM, SVSM, XRD and FEG-SEM facilities.